\documentclass[preprint,11pt]{article}

\usepackage{graphicx,epsf,amsbsy,amsmath,amssymb}
\usepackage[latin1]{inputenc}

\begin{document}

\title{Van der Pol and the history of relaxation oscillations: \\
toward the emergence of a concept}

\author{Jean-Marc Ginoux,\\ Universit\'e Pierre et Marie Curie, Paris VI,\\
Institut de Math\'ematiques de Jussieu, UMR 7586\\ ginoux@univ-tln.fr, http://ginoux.univ-tln.fr\\
Christophe Letellier\\ CORIA UMR 6614, Universit\'e de Rouen,\\ 
Saint-Etienne du Rouvray cedex, France}

\maketitle

\begin{abstract}
Relaxation oscillations are commonly associated with the name of Balthazar van
der Pol {\it via} his eponymous paper (\textit{Philosophical Magazine}, 1926)
in which he apparently introduced this terminology to describe the
nonlinear oscillations produced by self-sustained oscillating systems such as
a triode circuit. Our aim is to investigate how relaxation oscillations were
actually discovered. Browsing the literature from the late 19th century, we
identified four self-oscillating systems in which relaxation oscillations have
been observed: i) the series dynamo machine conducted by G\'erard-Lescuyer
(1880), ii) the musical arc discovered by Duddell (1901) and investigated by
Blondel (1905), iii) the triode invented by de Forest (1907) and, iv) the
multivibrator elaborated by Abraham and Bloch (1917). The differential equation
describing such a self-oscillating system was
proposed by Poincar\'e for the musical arc (1908), by Janet for the
series dynamo machine (1919), and by Blondel for the triode (1919). Once
Janet (1919) established that these three self-oscillating systems can be
described by the same equation, van der Pol proposed (1926) a generic
dimensionless equation which captures the relevant dynamical properties shared
by these systems. Van der Pol's contributions during the period of 1926-1930
were investigated to show how, with Le Corbeiller's help, he popularized the
``relaxation oscillations'' using the previous experiments as examples and,
turned them into a concept.
\end{abstract}

\maketitle

{\bf Van der Pol is well-known for the eponymous equation which describes a
simple self-oscillating triode circuit. This equation was used as a prototypical
model for many electro-technical devices where the triode is replaced with an
arc lamp, a glow discharge tube, an electronic vacuum tube, etc. If this
equation is now used with a driven term as a benchmark system producing chaotic
behavior, it was also associated with an important class of oscillations,
the so-called relaxation oscillations, where relaxations refers to the
discharge of a capacitor. Van der Pol popularized this name. But what was
his actual contribution? We bring some anwsers to this question.}

\section{Introduction}

Since the late 19th century, with the ability to conduct electro-mechanical
or electrical systems, sustained self-oscillating behaviors started to be
experimentally observed. Among systems able to present self-sustained
oscillations and which were investigated during this period, the triode circuit
is the most often quoted today, problably due to its relevant role played in
the development of the widely popular wireless telegraphy.   It is almost
impossible to find a reference to another system in the recent publications
and, van der Pol is widely associated with the very first development
of a nonlinear oscillations theory.

In the 60's, van der Pol was already considered as ``one of the most eminent
radio scientists, a leader in many fields, and one of the pioneers in the
theory and applications of nonlinear circuits'' \cite{Stu60}. Mary-Lucy
Cartwright, considered \cite{Car60} that ``his work formed the basis of much of
the modern theory of nonlinear oscillations. [...] In his paper \cite{VdP26},
on ``Relaxation-Oscillations'' van der Pol was the first to discuss
oscillations which are {\it not} nearly linear.'' After such a
statement by Cartwright, the first woman mathematician to be elected as a
Fellow of the Royal Society of England who worked with John Littlewood on
the solutions to the van der Pol equation \cite{Car45}, probably triggered the
historiography in this domain to be mainly focussed on van der Pol's
contribution and, particularly, on Ref. \cite{VdP26} in which he
``apparently'' introduced the term {\it relaxation} to characterize the
particular type of nonlinear oscillations he investigated.

The birth of relaxation oscillations is thus widely recognized as being
associated with this paper. Since these oscillations were seemingly discovered
by van der Pol as a solution to his equation describing a triode circuit,
historians looked for the first occurrence of this equation, and led to the
conclusion that it was written four years before 1926 according to Israel
\cite[p. 4]{Israel} :

\begin{quote}
``Of particular importance in our case is the 1922 publication in collaboration with Appleton, as it contained an embryonic form of the equation of the triode oscillator, now referred to as ``van der Pol's equation''.
\end{quote}
This was summarized in Aubin and Dahan-Dalmedico \cite[p. 289]{Aubin} as
follows.

\begin{quote}
``By simplifying the equation for the amplitude of an oscillating current
driven by a triode, van der Pol has exhibited an example of a dissipative
equation without forcing which exhibited sustained spontaneous oscillations:
\[
\nu'' - \varepsilon \left( 1 - \nu^2 \right) \nu' + \nu = 0
\]
In 1926, when he started to investigate its behaviour for large values of
$\varepsilon$ (where in fact the original technical problem required it to be
smaller than 1), van der Pol disclosed the theory of relaxation oscillation.''
\end{quote}
Dahan \cite[p. 237]{Dahan} goes even further by stating that
``van der Pol had already made use of Poincar\'e's limit cycles''.

Thus, according to the reconstruction proposed by these historians it seems
that van der Pol's contribution takes place at several levels: i) writing the
``van der Pol equation'', ii) discovering a new phenomenon named relaxation
oscillations and iii) recognizing that these oscillations were a limit cycle.
But concerning the last point, Diner \cite[p. 339]{Diner} with others
claimed that:

\begin{quote}
``Andronov recognizes, for the first time that in a radiophysical oscillator
such as van der Pol's one, which is a nonconservative (dissipative) system and
whose oscillations are maintained by drawing energy from non vibrating sources, the motion in the phase space is a kind of limit cycle, a concept introduced by
Poincar\'e in 1880 in a pure mathematical context.''
\end{quote}
Nevertheless, this historical presentation is not consistent with the facts
and provides a further illustration of the ``Matthew effect'' \cite{Merton}.
By focusing almost exclusively on some van der Pol's publications dealing with
a self-oscillating triode, the side effect was to partly or completely hide
previous research on sustained oscillations and to overestimate van der Pol's
contribution.

Our approach was thus to browse the literature from the last two decades of
the 19th century up to the 1930's to check whether sustained oscillations were
not investigated in some other mechanical or electrical systems. We did not
pretend to be exhaustive since our investigations remain to be improved,
particularly concerning the German literature. Nevertheless, we were
already able to find works preceding van der Pol's contributions, some of them
being even quoted in some of his papers. A more extended analysis of these
aspects can be found in \cite{Gith}. Section \ref{orig} is devoted to systems
producing sustained oscillations and which were published before 1920.
Section \ref{vdpeq} is focussed on how these experimental systems were
described in terms of differential equations and, on the origin of the
so-called ``van der Pol equation''. Section \ref{solu} discusses how
``relaxation oscillations'' were introduced and how they were identified in
various systems. Section \ref{conc} gives a conclusion.

\section{The early self-oscillating systems}
\label{orig}

\subsection{A series dynamo machine}

In 1880, during his research at the manufacture of electrical generators, the
French engineer Jean-Marie-Anatole G\'erard-Lescuyer made an experiment
where he coupled a dynamo acting as a generator to a magneto-electric machine.
G\'erard-Lescuyer \cite[p. 215]{GL2} observed a surprising phenomenon he
described as follows.

\begin{quote}
``As soon as the circuit is closed the magnetoelectrical machine begins to
move; it tends to take a regulated velocity in accordance with the intensity of
the current by which it is excited; but suddenly it slackens its speed, stops,
and start again in the opposite direction, to stop again and rotate in the same
direction as before. In a word, it receives a regular reciprocating motion
which lasts as long as the current that produces it.''
\end{quote}
He thus noted periodic reversals in the rotation of the magneto-electric
machine although the source current was constant. He considered this
phenomenon as an ``electrodynamical paradox''. According to him, an increase in
the velocity of
the magneto-electric machine gives rise to a current in the opposite direction
which reverses the polarity of the inductors and, then reverses its rotation.

In fact, many years later Paul Janet (1863-1937) \cite{Janet1,Janet2} revealed
that between brushes of the dynamo there is an electromotive force (or a
potential difference) which can be represented as a nonlinear function of the
current. Such a nonlinear current-voltage characteristic lead to an explanation
for the origin of this phenomenon.

\subsection{The musical arc}

\begin{figure}[ht]
  \begin{center}
    \includegraphics[height=4.3cm]{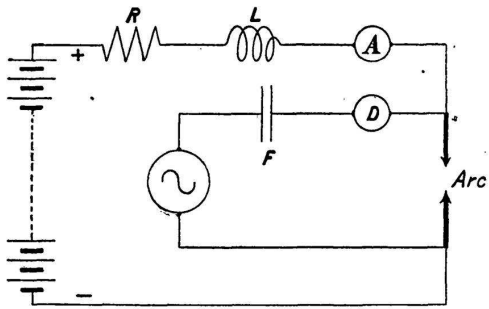} \\[-0.2cm]
    \caption{Diagram of the circuit built by Duddell to drive the electric
arc with an oscillating circuit \cite{Dud01}.}
    \label{dudstop}
  \end{center}
\end{figure}

By the end of the 19th century an electric arc was used for lighthouses and
street lights.
Regardless of its weak light, it had a major drawback: the noise produced by
the electrical discharge was disturbing the residents. In London,
William Du Bois Duddell (1872-1917) was commissioned in 1899 by the
British authorities to solve this problem. He stopped the rustling by inserting
a high-frequency oscillating circuit in the electric arc circuit (Fig.\
\ref{dudstop}): as he wrote \cite[p. 238]{Dud01}, ``the sound only became
inaudible at frequencies approaching 30,000 oscillations per second.''

\begin{figure}[ht]
  \begin{center}
    \includegraphics[height=4.2cm]{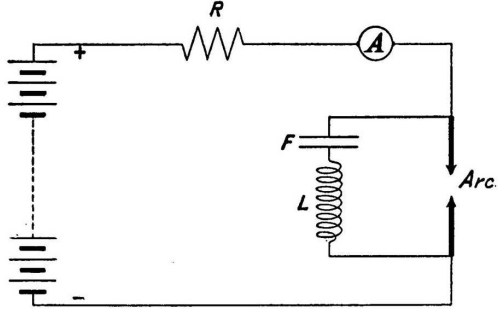} \\[-0.2cm]
    \caption{Diagram of the circuit with a constant source of current built by
Duddell to obtain a musical arc \cite{Dud01}.}
    \label{musicarc}
  \end{center}
\end{figure}

During his investigations, Duddell obtained other interesting results. He also
built an oscillating circuit with an electrical arc and a
constant source \cite{Dud07}:

\begin{quote}
``... if I connect between the electrodes of a direct current arc [...] a
condenser and a self-induction connected in series, I obtain in this shunt
circuit an alternating current. I called this phenomenon the {\it musical}
arc. The frequency of the alternating current obtained in this shunt circuit
depends on the value of the self-induction and the capacity of the condenser,
and may practically be calculated by Kelvin's well-known formula \cite{Tho}.''
\end{quote}
The block diagram of this circuit is shown in Fig. \ref{musicarc}.

Duddel well identified that it is

\begin{quote}
``the arc itself which is acting as a converter and transforming a part of the
direct current into an alternating, the frequency of which can be varied
between very wide limits by alterning the self-induction and capacity.''
\end{quote}
To observe this effect, Duddell showed that the change in the potentiel
$\delta V$ and in the corresponding current $\delta A$ between the ends
of the arc must obey to $-\frac{\delta V}{\delta A} \geq r$ where $r$ is the
resistance of the capacitor circuit. This means that the arc can be seen
as presenting a negative resistance (a concept introduced by Luggin
\cite{Luggin}). The various behaviors obtained by Duddell were as follows.

\begin{figure}[ht]
  \begin{center}
    \begin{tabular}{c}
      \includegraphics[height=3.3cm]{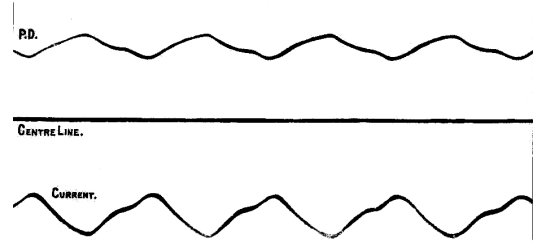} \\
      (a) Humming: $\overline{U} = 50.5$ V, $\overline{I} = 15.2$ A \\
      Top curve = potential difference (PD) \\ Bottom curve = current \\[0.2cm]
      \includegraphics[height=3.5cm]{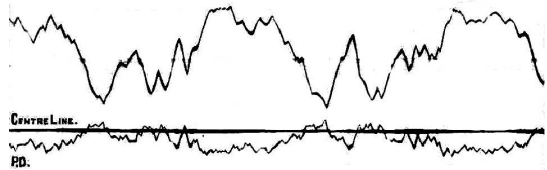} \\
      (b) Hissing: $\overline{U} = 38$ V, $\overline{I} = 22.3$ A \\
      Top curve = current \\ Bottom curve = potential difference (PD)
      \\[-0.2cm]
    \end{tabular}
    \caption{Two types of oscillating arc supplied with an oscillating current.
Centre line = 40 V = 20 A. With the original figure size, 1 mm = 0.5 V = 0.186
A = $\frac{1}{6400} s$. From \cite{Dud01}.}
    \label{dudbehav}
  \end{center}
\end{figure}

\begin{itemize}

\item[i)] When the arc is supplied with an oscillating current, he obtained
{\it humming} which are nearly periodic (Fig.\ \ref{dudbehav}a) and {\it
hissing} which are very irregular and whose frequence is such as
$f_{\rm hissing} \approx 100 f_{\rm humming}$ (Fig.\ \ref{dudbehav}b).

\item[ii)] When the arc is supplied with a constant current, he obtained a
self-oscillating electrical arc, a regime he called intermittent or musical.

\end{itemize}

Hissing and intermittent regimes clearly correspond to nonlinear oscillations
since not obeying Thomson's formulae for the period of oscillations.
Duddell proposed the formula
\begin{equation}
  T = \frac{2 \pi}{\displaystyle
      \sqrt{\displaystyle \frac{1}{LC} - \frac{R^2}{4L^2}} }
\end{equation}
for their period \cite{Dud01c}.
But there is an ambiguity in Duddell's work since intermittent (non-linear)
and musical (linear since obeying to Thomson's formula) are not distinguished.
Duddell quoted Andr\'e Blondel (1863-1938) for his observation of the hissing
and the intermittent arcs \cite{Blo92}.

\begin{figure}[ht]
  \begin{center}
    \begin{tabular}{c}
      \includegraphics[height=4.8cm]{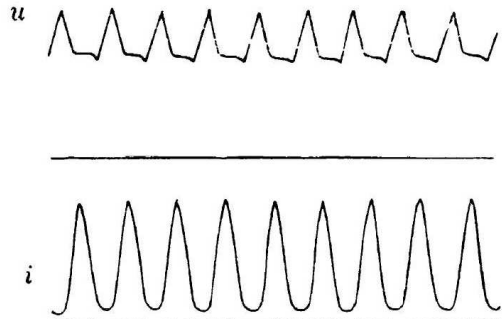} \\
      (a) Musical oscillations \\[0.2cm]
      \includegraphics[height=4.0cm]{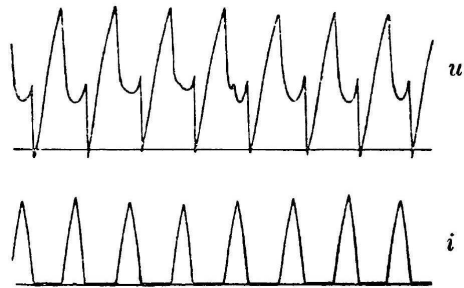} \\
      (b) Discontinuous or intermittent oscillations \\[-0.2cm]
    \end{tabular}
    \caption{The two types of oscillations produced by a musical arc and
distinguished by Blondel. Musical oscillations are very stable but
discontinuous (intermittent) oscillations are only observed when the arc is
close to its limit of stability. From \cite{Blo05}.}
    \label{blondosc}
  \end{center}
\end{figure}

Blondel was assigned to the Lighthouses and Beacons Service, to conduct
research on electric arc in order to improve the outdated French electrical
system. He made an extensive analysis of the musical arc \cite{Blo05} using
Duddell's circuit (Fig.\ \ref{musicarc}). Blondel, who named it the ``singing
arc'', clearly distinguished two types of oscillations which were not very
well defined in Duddell's work. The first type corresponds to musical
oscillations as used by Duddell to play tunes at a meeting of the Institution
of Electrical Engineers using an arrangement of well-chosen capacitors and a
keyboard \cite{Dud01}. According to Blondel, these oscillations are nearly
sinusoidal (Fig.\ \ref{blondosc}a). The second type corresponds to
discontinuous oscillations (Fig.\ \ref{blondosc}b) and were termed
``intermittent''
by Duddell. In these regimes, already observed in 1892 by Blondel \cite{Blo92}
and quoted in \cite{Dud01}, the arc can blow out and relight itself with great
rapidity. Blondel explained that such intermittent (or discontinuous) regimes
are observed when the circuit is placed close to the limit of stability of the
arc. Unfortunately, Blondel qualified these second type of oscillations as
hissing (``sifflantes'' or ``stridentes'') although they have nothing o do
with the hissing regime observed by Duddell (compare Fig.\ \ref{dudbehav}b with
Fig.\ \ref{blondosc}b).

\begin{figure}[ht]
  \begin{center}
    \includegraphics[height=3.8cm]{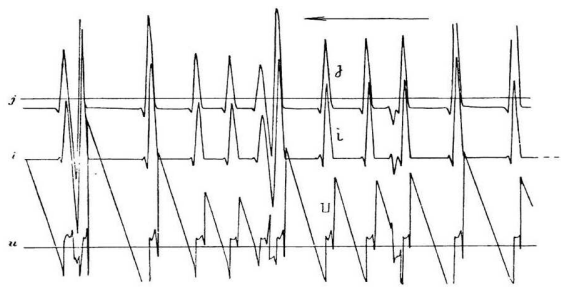} \\[-0.2cm]
    \caption{Oscillations ``with 3 or 4 alternations'' observed with an
musical arc between a copper anode and a carbon cathod. From \cite{Blo05}.}
    \label{intirreg}
  \end{center}
\end{figure}

Blondel also observed irregular intermittent  oscillations as shown in Fig.\
\ref{intirreg}.  These periodic oscillations are quite interesting today
because they have obviously recurrent properties and could therefore be a good
candidate for being considered as chaotic.

\subsection{From the Audion to the Multivibrator}

\begin{figure}[htbp]
  \begin{center}
    \includegraphics[height=5.5cm]{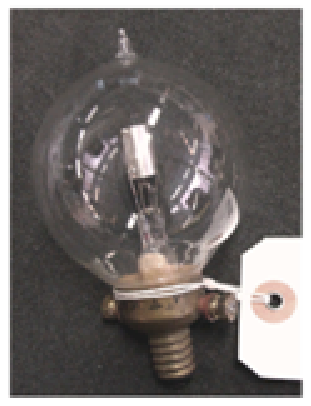} \\[-0.2cm]
    \caption{Picture of the original lamp T.M. made by Abraham (1915).}
    \label{Triode}
  \end{center}
%
  \begin{center}
    \begin{tabular}{c}
      \includegraphics[width=7.35cm,height=6.5cm]{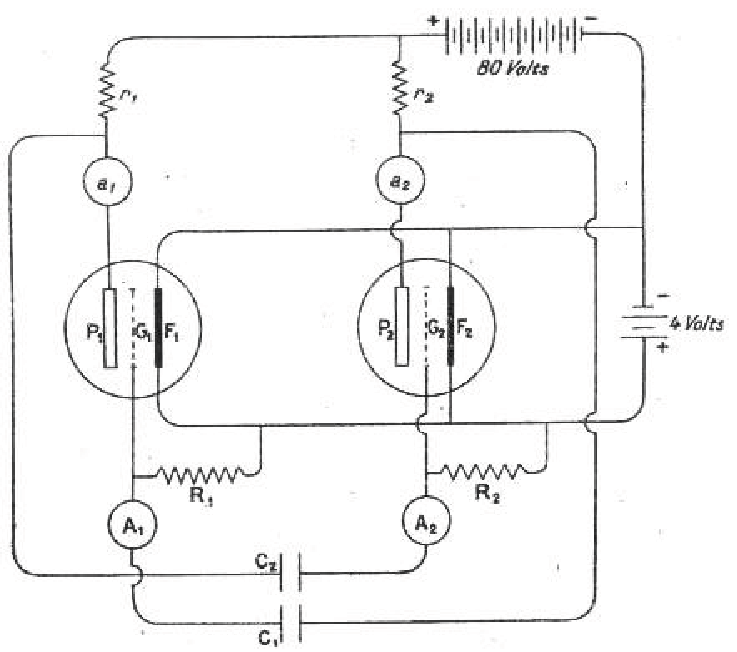} \\
      (a) Circuit diagram of multivibrator. \\[0.3cm]
      \includegraphics[width=7.0cm]{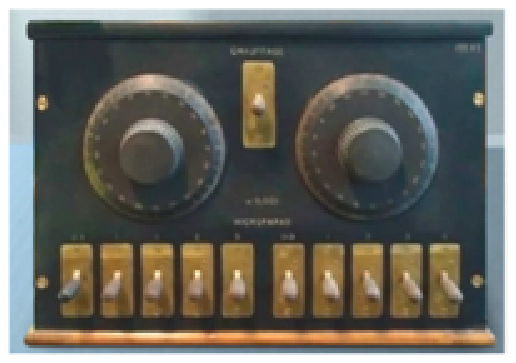} \\[-0.0cm]
      (b) Picture of the original multivibrator. \\[-0.2cm]
    \end{tabular}
    \caption{Circuit diagrams and picture of the multivibrator.}
    \label{Mu1}
  \end{center}
\end{figure}

Although the ``audion'', a certain type of triode, was invented in 1907 by
Lee de Forest (1873-1961) \cite{For07}, it was not until the First World War
that it was widely distributed for military and commercial purposes. During the
summer of 1914, the French engineer Paul Pichon (18??-1929) who deserted the
French Army in 1900 and had emigrated to Germany, went to USA on an assignment
from his employers, the Telefunken Company of Germany, to gather samples of
recent wireless equipments. During his tour he visited
the Western Electric Company, and obtained the latest high-vacuum Audion
with full information on their use. On his way back to Germany
he traveled by the Atlantic liner to Southampton. He arrived in London on
August 3, 1914, the very day upon which Germany declared war on France.
Considered as a deserter in France and as an alien in Germany, he decided to go to Calais where he was arrested and brought to the French military authorities
which were represented by Colonel Gustave Ferri\'e (1868-1932), the commandant
of the French Military Telegraphic Service. Ferri\'e immediately submitted
Pichon's Audion to eminent
physicists including Henri Abraham (1868-1943) who was then sent to Lyon in
order to reproduce and improve the device. Less than one year after, the
French valve known as the lamp T.M. (Military Telegraphy) was born (Fig.
\ref{Triode}). After many tests it became evident that the French valve was
vastly superior in every way to the soft-vacuum Round valves and earlier
Audions.

On May 1915, Ferri\'e asked to Abraham to come back at the \'Ecole Normale
Sup\'erieure (Paris) where, with his colleague Eugène Bloch (1878-1944), he
invented the ``multivibrator'' during the first World War. A multivibrator is a
circuit (Fig.\ \ref{Mu1}) made of two lamps T.M. producing sustained
oscillations with many harmonics.

\begin{figure}[ht]
  \begin{center}
    \includegraphics[height=3.00cm]{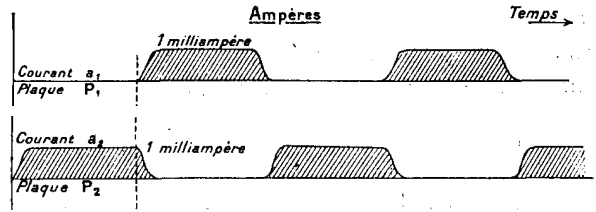} \\[-0.4cm]
    \caption{Reversals of the current in plate P$_{1}$ and P$_{2}$ currents.
From \cite[p. 256]{Abr19}.}
    \label{Mu2}
  \end{center}
\end{figure}

Made of two resistors ($R_1$ and $R_2$) and two capacitors ($C_1$ and $C_2$),
Abraham and Bloch \cite[p. 256]{Abr19} explained that reversal of current
intensities in plates $P_1$ and $P_2$ were observed (Fig. \ref{Mu2}). They
described these oscillations as ``a serie of very sudden reversals in currents
splitted by long intervals during which variation of the current intensity is
very slow'' \cite{Abr19}. Two reversals
were separated by time intervals corresponding to durations of charge and
discharge of capacitors $C_1$ and $C_2$ through resistors $R_1$ and $R_2$,
respectively. ``The period of the system was thus about $C_1 R_1 +C_2 R_2$.''
The period of these oscillations is therefore associated with some capacitor
discharge.

\section{Differential equations for self-oscillating systems}
\label{vdpeq}

\subsection{Poincar\'e's equation for the musical arc}

In 1880's, Poincar\'e developed his mathematical theory for differential
equations and introduced the concept of limit cycle \cite{P4} as

\begin{quote}
``closed curves which satisfy our differential equations and
which are asymptotically approached by other curves defined by the same
equation but without reaching them.''
\end{quote}
Aleksandr Andronov (1901-1952) was commonly credited for the first evidence of
a limit cycle in an applied problem, namely in self-sustained oscillating
electrical circuit \cite{Andro1}. Nevertheless, it was recently found by one
of us \cite{Gin10} that Poincar\'e gave a series of lectures at the
\'Ecole Sup\'erieure des Postes et T\'el\'egraphes (today Sup'T\'el\'ecom) in
which he established that the existence of sustained oscillations in a musical
arc corresponds to a \textit{limit cycle} \cite{Poi08}.

\begin{figure}[ht]
  \begin{center}
    \includegraphics[height=3.0cm]{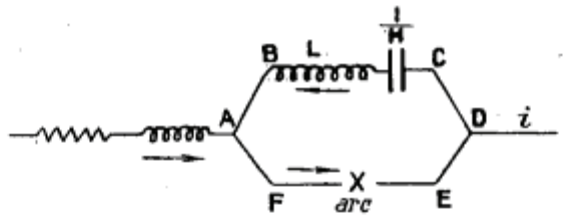} \\[-0.4cm]
    \caption{Block diagram of the circuit corresponding to the musical arc
as investigated by Poincar\'e in 1908 \cite[p. 390]{Poi08}.}
    \label{Po1}
  \end{center}
\end{figure}

Investigating Duddell's circuit (Fig.\ \ref{Po1} which is equivalent to
Fig.\ \ref{musicarc}), Poincar\'e wrote the corresponding differential
equations. Designating the capacitor charge by $x$, the
current in branch ABCD was thus $x' = \frac{dx}{dt}$.
Applying Kirchhoff's law, Poincar\'e obtained the second-order nonlinear
differential equation  \cite{Poi08}
\begin{equation}
  \label{Poin1}
  Lx'' + \rho x' + \theta \left( x' \right) + H x = 0 \, .
\end{equation}
This equation has a general form since the circuit characteristic $\theta
\left( x' \right)$ describing how the electromotive force
depends on the current was not explicitly defined. $\rho x'$ corresponds to the
internal resistance of the inductor and to the other possible causes of damping,
including the radiation from the antenna. $\theta (x')$ is the term due to
the arc and $\frac{1}{H}$ defines the capacitor. Using the coordinate
transformation
\begin{equation}
  x' = \frac{dx}{dt} = y \, , \, dt = \frac{dx}{dy} = y \, , \,
  x'' = \frac{dy}{dt} = \frac{ydy}{dx} \, ,
\end{equation}
introduced in \cite[p. 168]{P4}, Poincar\'e obtained
\begin{equation}
  \label{Poin2}
  Ly\frac{dy}{dx}  + \rho y + \theta \left( y \right) + H x = 0 \, .
\end{equation}
He then stated that

\begin{quote}
``One can construct curves satisfying this differential equation, provided
that function $\theta $ is known. Sustained oscillations correspond to closed
curves, if there exist any. But every closed curve is not appropriate, it must
fulfill certain conditions of stability that we will investigate.''
\end{quote}
Once the direction of rotation defined, Poincar\'e stated the stability
condition \cite[p. 391]{Poi08}:

\begin{quote}
``Let us consider another non-closed curve satisfying the differential
equation, it will be a kind of spiral curve indefinitely approaching the closed
curve. If the closed curve represents a stable regime, by describing the
spiral in the direction of the arrow, one should be brought to the closed
curve, and this is the condition according to which the closed curve will
represent a stable regime of sustained waves and will provide a solution to
this problem.''
\end{quote}
The stability condition obviously matches with the definition of a limit cycle.
In order to define under which condition a closed curve is stable, Poincar\'e
multiplied Eq. (\ref{Poin2}) by $x' {\rm d}t$ and then integrated the obtained
relation over one period of oscillation. Using the fact that the first
and the fourth term vanish since corresponding to the conservative component
of his nonlinear equation, the obtained condition reduces to
\begin{equation}
  \label{Poin3}
  \rho \int {{x}'^2dt} +\int {\theta \left( {{x}'} \right){x}'dt} = 0 \, .
\end{equation}
Since the first term is quadratic, the second one must be negative. Poincar\'e
then stated that the oscillating regime is
stable if and only if function $\theta$ is such as
\begin{equation}
  \label{Poin4}
  \int {\theta \left( {{x}'} \right){x}'dt} < 0 \, .
\end{equation}
Poincar\'e thus provided a condition the characteristic $\theta (x')$ of the
circuit must fulfill to guaranty the presence of a limit cycle in a musical
arc. Using the Green formula, this condition was proved
\cite{Gin10} to be equivalent to the condition obtained by Andronov
\cite{Andro1} more than twenty years later. It is therefore relevant to realize
that Poincar\'e was in fact the first to show that his ``mathematical'' limit
cycle was of importance for radio-engineering. Up to now Andronov was therefore
erroneoulsy credited for such insight with a more general equation in
his 1929.

\subsection{Janet's equation for series dynamo machine}

In April 1919, Janet exhibited an analogy between three electro-mechanical
devices, namely i) the series dynamo machine, ii) the musical arc and
iii) the triode. For the series dynamo machine, Janet quoted Aim\'e
Witz (1848-1926) \cite{Wit89} who was presenting the experiment as widely
known. He only quoted G\'erard-Lescuyer in his introduction to Cartan and
Cartan's paper \cite{Ca}. Janet \cite[p. 764]{Janet1} wrote about the series
dynamo machine:

\begin{quote}
``It seemed interesting to me to report unexpected analogies between this
experiment and the sustained oscillations so widely used today in wireless
telegraphy, for instance, those produced by Duddell's arc or by the
three-electrodes lamps used as oscillators. The production and the maintenance
of oscillations in all these systems mainly result from the presence, in the
oscillating circuit, of something analogous to a negative resistance.''
\end{quote}
He then explained that function $e = f(i)$, where $e$ is the potential
difference and $i$ the current, should be nonlinear and should act as a
negative resistance. He thus established that these three devices produced
non-sinusoidal oscillations and could be described by a unique nonlinear
differential equation reading as \cite[p. 765]{Janet1}
\begin{equation}
  \label{Jan1}
  L\frac{d^2i}{dt^2} + \left[ {R-{f}'\left( i \right)} \right]\frac{di}{dt}
  +\frac{k^2}{K}i=0
\end{equation}
where $R$ corresponds to the resistance of the series dynamo machine,
$L$ is the self-induction of the circuit and $K/k^2$ is analogous to a
capacitor.

Replacing $i$ with $x$, $R$ with $\rho$, $f'(i)$ with $\theta (x)$, and
$\frac{k^2}{K}$ with $H$, this equation becomes Poincar\'e's equation
(\ref{Poin1}). As in Poincar\'e's work, Janet did not provide an explicit
form for nonlinear function $f(i)$. Nevertheless,  he wrote the oscillating
period as being $T= \frac{2 \pi}{k} \sqrt{KL}$, which is equivalent to
$T = 2 \pi \sqrt{CL}$ as obtained by Duddell when resistance $R$ is neglected.

Duddell was quoted by Janet
without reference. Janet spoke about Duddell's arc as a very well-known
experiment. Being an engineer in electro-technique, Janet was a reader of {\it
L'\'Eclairage \'Electrique} in which he published more than 20 papers.
Consequently, he may have read Poincar\'e's conferences...

\subsection{Blondel's equation for the triode}

\begin{figure}[ht]
  \begin{center}
    \begin{tabular}{cc}
      \includegraphics[height=3.6cm]{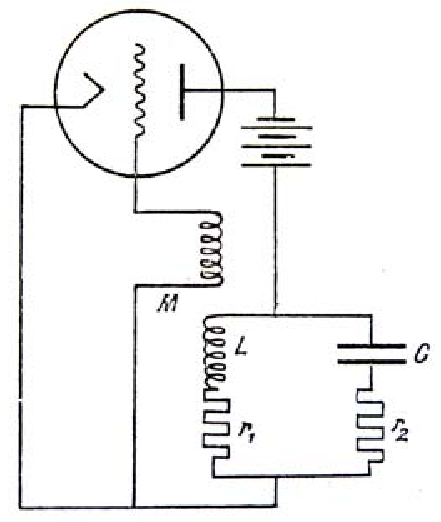} &
      \includegraphics[height=3.6cm]{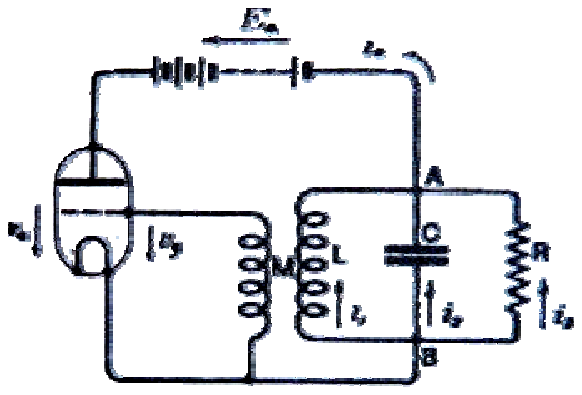} \\[-0.0cm]
      (a) From Blondel \cite{B2} & (b) From Van der Pol \cite{VdP1} \\[-0.2cm]
    \end{tabular}
    \caption{Circuit diagrams for the triode oscillator as investigated by
Blondel (a) and van der Pol (b). Resistance $r_2$ in Blondel's circuit does not
appear in van der Pol's circuit where resistance $R$ was added in parallel to
capacitor $C$.}
    \label{Bo2}
  \end{center}
\end{figure}

The ``triode'' equations was proposed six months later by Blondel who used a
circuit with an Audion. To achieve this, he proposed to approximate the
characteristic as a series of odd terms \cite[p. 946]{B2}
\begin{equation}
  \label{eqB1}
     i = b_1 \left(u + k \nu \right)
      - b_3 \left(u + k \nu \right)^3 - b_5 \left(u + k \nu \right)^5 ...
\end{equation}
From the circuit diagram he used (Fig. \ref{Bo2}a) and applying Kirchhoff's
law, he got
\begin{equation}
  \label{eqB2}
  \begin{array}{rl}
    \displaystyle
    \frac{{\rm d}^3u}{{\rm d}t^3} + \frac{r_2}{L}\frac{{\rm d}^2u}{{\rm d}t^2}
    + \left(\frac{1}{CL}
    - \frac{r_1r_2}{L^2} \right)\frac{{\rm d}u}{{\rm d}t} & \\[0.3cm]
    \displaystyle
    - \frac{r_1}{CL^2}u - r_2 \frac{{\rm d}^3i}{{\rm d}t^3}
    - \frac{1}{C}\frac{{\rm d}^2i}{{\rm d}t^2}
   & = 0
  \end{array}
\end{equation}
By substituting $i$ by its expression in Eq. (\ref{eqB2}), neglecting the
internal resistors and integrating once with respect to time, he thus
obtained
\begin{equation}
  \label{eqB3}
  C \frac{{\rm d}^2u}{{\rm d}t^2} - \left(b_1h - 3b_3 h^3 u^2
  - ... \right)\frac{{\rm d}u}{{\rm d}t} + \frac{u}{L} = 0 \, ,
\end{equation}
that is, an equation equivalent to those obtained by Poincar\'e and Janet,
respectively, but where the characteristic was explicitely written. Blondel
considered this equation as an approximation.

\subsection{The van der Pol equation}

One year after Blondel's paper, van der Pol \cite{VdP1} proposed an equation
for a triode oscillator. Its circuit diagram (Fig. \ref{Bo2}b) differs from
Blondel's one by the internal resistance he neglected and a resistance
$R$ he added in parallel to the capacitor. Van der Pol used a Taylor-McLaurin
series expansion to express the characteristic of the triode by the cubic
function \cite[p. 703]{VdP1}
\begin{equation}
\label{eqV1}
i = \psi\left(k v \right) = \alpha v + \beta v^2 + \gamma v^3 \, .
\end{equation}
He then obtained the differential equation
\begin{equation}
  \label{eqV2}
  C \frac{{\rm d}^2 v}{{\rm d}t^2}
  + \left(\frac{1}{R} - \alpha \right)
  \frac{{\rm d}v}{{\rm d}t}
  + \beta \frac{{\rm d} \nu^2 }{{\rm d}t}
  + \gamma \frac{{\rm d} \nu^3 }{{\rm d}t} = 0
\end{equation}
Van der Pol \cite[p. 704]{VdP1} precised that, by symmetry consideration, one
can choose $\beta = 0$. However, to allow comparison with Blondel's equation
(\ref{eqB3}), one should also remove the resistance $R$. With Appleton, van der
Pol then reduced Eq. (\ref{eqV2}) \cite{App22} to
\begin{equation}
  \label{eqV3}
  \frac{{\rm d}^2 v}{{\rm d}t^2} + \frac{{\rm d}}{{\rm d}t}
  \left( \displaystyle \frac{Rv}{L} + \frac{\Psi (v)}{C} \right)
  + \omega_0^2 v = 0
\end{equation}
where
\begin{equation}
  \Psi (v) = \alpha v + \beta v^2 + \gamma v^3 + \delta v^4 + \epsilon v^5 +
  ...
\end{equation}
They also provided a stability criterion for periodic solutions when
coefficients $\alpha$, $\beta$, $\gamma$, etc. are sufficiently small.
Solutions to Eq. (\ref{eqV3}) were then investigated by Appleton and Greaves
(see for instance \cite{App22b,App23}). This is only in 1926 that van der Pol
introduced the dimensionless equation
\cite{VdP4,VdP5}
\begin{equation}
  \label{protovdp}
  \ddot{v} - \epsilon (1 - v^2) \dot{v} + v = 0 \, .
\end{equation}
This equation quicly became popular in radio-electricity. It can be found in a
book review (1935) by Robert Mesmy \cite{Mes35} associated with van der Pol's
name and the relaxation oscillations.  In
\cite[p. 373]{Andro2}, this equation was mentioned as follows.

\begin{quote}
{\small ``As a typical example of the technique of isoclines, one can use the
study of an equation proposed by van der Pol in the phase plane which, we
already know it, is called the van der Pol equation''}
\end{quote}
Quoting \cite{Andro2}, Minorsky \cite{Min47} used the same name. If the van der
Pol equation is only an example among many others in Ref. \cite{Andro2}, its
role is far more important in \cite{Min47}. Van der Pol's breakthrough was to
propose a
dimensionless equation which can thus be used to explain various systems
regardless their origins. From that point of view, van der Pol was the first to
write a propotypal equation. In doing that, he assimilated the fact that the
dynamical nature of the behaviours observed was much more important than a
physical explanation as provided, for instance, by Blondel \cite{Blo05,B2}.
Van der Pol is therefore correctly credited for this equation written in its
simplest form in 1926. Van der Pol also pushed the analysis further by
investigating some solutions to his equation.

\subsection{Some equations for the multivibrator}

In his 1926 paper \cite{VdP26}, van der Pol proposed an equation for the
multivibrator based
on current and potential departures from the unstable equilibrium
values. Surprisingly, he had to take into account ``the inductance $L$ of the
wires connected to the two capacities'' but neglected, ``for simplicity, the
influence of the anode potential on the anode current.'' Assuming that the
triodes are exactly equal, van der Pol showed that the prototypical equation
(\ref{protovdp}) ``represents the action of the multivibrator''. Consequently,
``the special vibration of the multivibrator represents an example of a
general type of relaxation-oscillations''.

\begin{figure}[ht]
  \begin{center}
    \includegraphics[height=4.5cm]{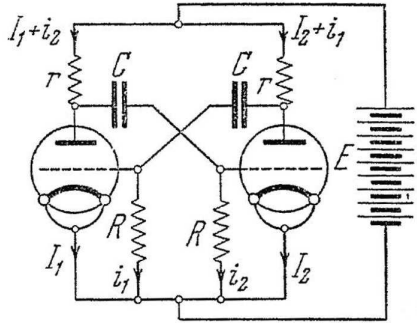} \\[-0.3cm]
    \caption{Block diagram of the multivibrator considered by Andronov and
Witt (from \cite{And30}).}
    \label{andcirc}
  \end{center}
\end{figure}

Without taking into account the inductance of the wires, Andronov and Witt
obtained another set of equations \cite{And30}
\begin{equation}
  \label{andeq30}
  \left\{
    \begin{array}{l}
      \displaystyle
      \dot{x} = \frac{ax - y \varphi' (y)}{\phi' (x) \varphi' (y) - a^2}
        \\[0.3cm]
      \displaystyle
      \dot{y} = \frac{ay - x \varphi' (x)}{\phi' (x) \varphi' (y) - a^2}
    \end{array}
  \right.
\end{equation}
where $x = Ri$, $y=Ri$, and $a = \frac{r + R}{Rr}$ (Fig.\ \ref{andcirc}). The
characteristic $I_k = \varphi (Ri_k)$ ($k=1,2$) were approximated by
$\varphi (\xi) = \frac{1}{1 + \xi^2}$. Andronov and Witt performed a dynamical
analysis of their system of equations for the multivibrator.

\begin{figure}[ht]
  \begin{center}
    \includegraphics[height=7.5cm]{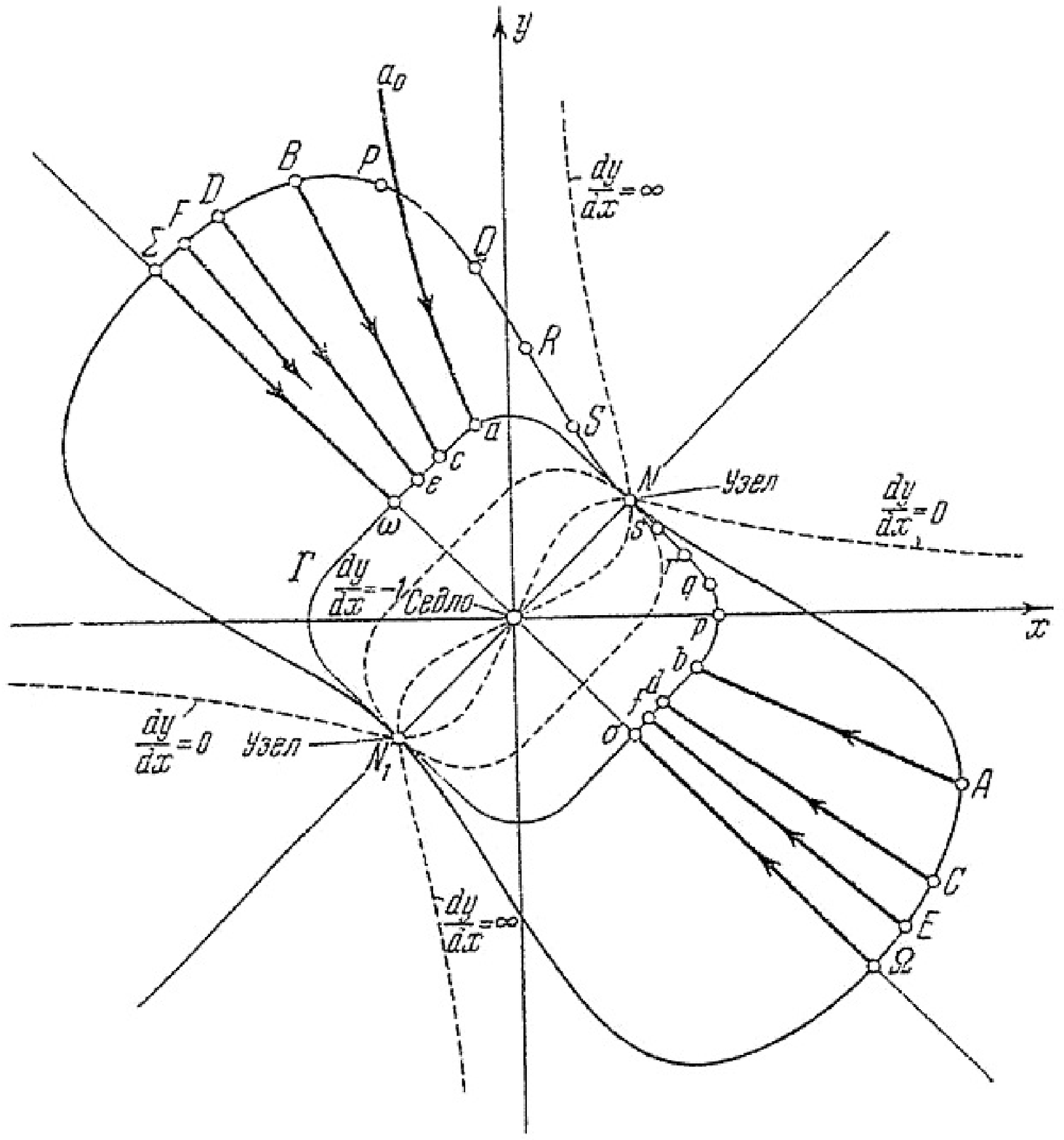} \\[-0.3cm]
    \caption{The trajectory solution to system (\ref{andeq30}) investigated by
Andronov and Witt (from \cite{And30}).}
    \label{andplane}
  \end{center}
\end{figure}

They found that
system (\ref{andeq30}) has three singular points, one located at the origin
of the phase space is a saddle and, two symmetry-related singular points
defined by
\begin{equation}
 x = y = \pm \sqrt{\displaystyle \frac{1 - a}{a} }
\end{equation}
are unstable (the type is not given with more precision). They then described
a typical trajectory (Fig.\ \ref{andplane}) starting from the initial condition
$a_0$. They observed a transient regime aAbBcCdDeEfF up to a $\omega$-limit
set, made of $\omega \Omega \sigma \Sigma$, which is a limit cycle. They
mentioned a ``discontinuous curve'' for which the jumps occur between $\omega$
and $\Omega$ and, $\sigma$ and $\Sigma$. This trajectory has the very
characteristics of ``relaxation oscillations'' but was not designated by such
a name.

\section{Birth of relaxation oscillations}
\label{solu}

Once the equation written, the job is just beginning. It remains to
investigate the
corresponding solutions. In 1925, van der Pol distinguished a ``separate
group'' of oscillations whose period can no longer be described by the
Thomson's fomula \cite{Tho}. Duddell, and then Blondel, already remarked that
``the oscillation frequency of the musical arc was variable and not well
defined [...] In the intermittent case, it has nothing to do with the
eigen-frequency of the oscillating circuit'' \cite{Blo05}. In 1925, van der Pol
remarked \cite{vdP25} that there are some oscillations which

\begin{quote}
{\small ``diverge considerably from the
sine curve, being more peaked. Moreover, the frequency is no longer determined
by the equation $\omega = \frac{1}{\sqrt{LC}}$ and the time period is
approximately the product of a resistance by a capacitor, i.e. a relaxation
time. Such oscillations belong to a separate group and are conveniently named
``relaxation-oscillations''.}''
\end{quote}
Van der Pol later noted that ``the period of these new oscillations is
determined by the duration of a capacitor discharge, which is sometimes named
a ``relaxation time'' \cite[p. 370]{VdP8}.
Since non-corresponding to the mostly investigated solutions, van der Pol
considered his results important enough to publish them at least four times:

\begin{enumerate}

\item Over Relaxatietrillingen \cite{VdP3} (in Dutch);

\item Over Relaxatie-trillingen \cite{VdP4} (in Dutch);

\item \"{U}ber Relaxationsschwingungen \cite{VdP5} (in German);

\item On relaxation-oscillations \cite{VdP26} (in English).

\end{enumerate}
All these contributions have the same title and almost the same content.
Consequently, the paper (in Dutch) which was published in 1925 should be quoted
as the first paper where relaxation oscillations were introduced.
Nevertheless, their conclusions differ in the choice for the devices
exemplifying the phenomenon of relaxation oscillations. They were

\begin{figure}[ht]
  \begin{center}
    \begin{tabular}{c}
      \includegraphics[height=3.5cm]{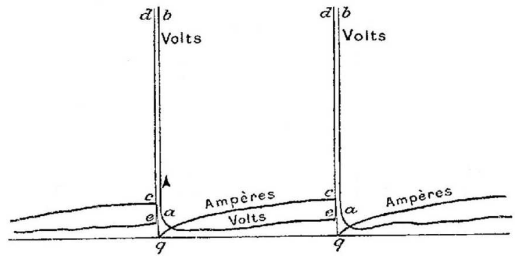} \\
      (a) From Blondel \cite{Blo99} \\[0.2cm]
      \includegraphics[height=2.8cm]{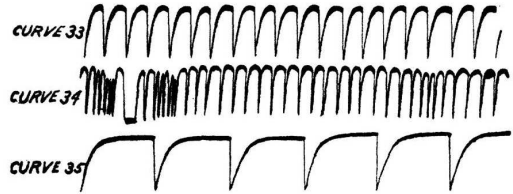} \\
      (b) From Compton \cite{Com10} \\[-0.2cm]
    \end{tabular}
    \caption{Current observed in a Wehnelt interrupter.}
    \label{wehnelt}
  \end{center}
\end{figure}

\begin{enumerate}

\item a Wehnelt interrupter (without reference): this example comes from
Blondel's paper \cite{Blo99} (Fig.\ \ref{wehnelt}a) but very suggestive
relaxation oscillations can also be found in \cite{Com10} (Fig.\
\ref{wehnelt}b);

\item a separate excited motor fed by a constant rotating speed series dynamo
and possibly heartbeats;

\item the three previous examples;

\item Abraham and Bloch's multivibrator,

\end{enumerate}
respectively. Van der Pol was therefore aware of these prior works presenting
self-sustained oscillations and whose some of them (Figs.\ \ref{dudbehav}b and
\ref{blondosc}b) were obviously nonlinear since not satisfying the Thomson
formula. Van der Pol recognized something special in these dynamics. In order
to enlarge the interest for these new dynamical features, he made an explicit
connection with electro-chemical systems (the Wehnelt interrupter). He may have
not quoted the electric arc because it became an obsolete device since Audion
was widely sold. More
surprisingly, van der Pol only quoted the well-known multivibrator in his most
celebrated paper \cite{VdP26}, which is the single one written in English (we
recall that not mentionning the first three examples is one of the very few
changes compared to his three previous papers). This could be explained by the
fact that only the multivibrator was then considered as a ``modern'' system.

The key point is that van der Pol started to investigate solution,
not when $\epsilon \ll 1$ as commonly considered, but when $\epsilon \gg 1$.
Van der pol distinguished various cases, speaking about the ``well-known
aperiodic case'', which refers to Pierre Curie's classification of transient
regimes \cite{Cur91}. Note that Eq. (1) in \cite{VdP26} is equivalent to Eq.
(2) in Curie's paper. Moreover, it is possible that van der Pol got the idea
to use a prototypical equation from Curie's paper where the study is performed
on a ``reduced equation''. This is in fact a dimensionless equation where
``each of its terms [...] has a null dimension with respect to the
fundamental units.'' Curie's paper was known from the electrical engineers
since it was quoted by Le Corbeiller in \cite{LeC31} who confirmed that ``M.
van der Pol [applied] it to the much more complicated case of Eq.
(\ref{protovdp}).'' Curie's paper was also used in \cite{Mes35} without
quotation and in \cite{Kry37} with an explicit quotation.

\begin{figure}[ht]
  \begin{center}
    \includegraphics[height=6.5cm]{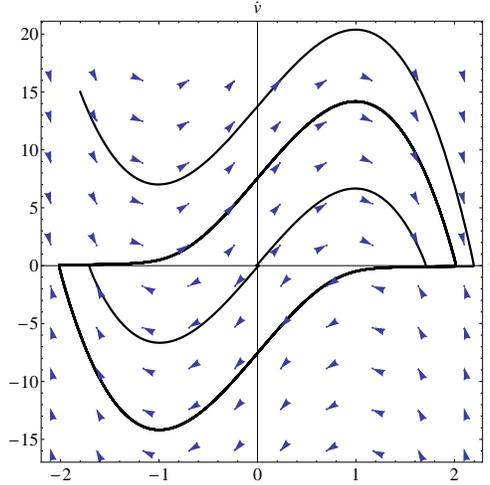} \\[-0.3cm]
    \caption{Isoclines in a $\dot{v}$-$v$ plane diagram. The heavy line
follow the slopes indicated on the isoclines, thus forming a solution to
equation (\ref{protovdp}). Drawn in the spirit of the original figure.}
    \label{solnull}
  \end{center}
\end{figure}

Using a series of isoclynes, that is, of curves connecting all points for which
$\frac{{\rm d}v}{{\rm d}t}$ is a given constant, van der Pol draw three cases
of oscillation for $\epsilon =0.1$, 1 and 10, respectively. He focussed his
attention on the third case (Fig.\ \ref{solnull})
which was designated as ``quasi-aperiodic'' because the oscillations
quickly converge toward a closed curve, the aperiodicity being only observed
during the transient regime. It actually seems that van der Pol erroneously
qualified the closed curve itself as an aperiodic behavior. It would be
an example of a lack of rigor and, source of contradiction as exemplified
in \cite[p. 987]{VdP26}: ``our equation for the quasi-aperiodic case, which
differs considerably from the normal approximately sinusoidal
solution, has again a {\it purely periodic} solution, ...'' How can we
have a clear understanding of a {\it quasi-aperiodic} solution which is {\it
purely periodic}?

The plot of the periodic solutions of equation (\ref{protovdp}) was one of the
result which was mostly appreciated in van der Pol's paper. Van der Pol did not
realized that this periodic solution was Poincar\'e's limit cycle. He did not
use this expression before 1930 \cite[p. 294]{VdP9}

\begin{quote}
``We see on each of these three figures a closed integral curve; this is an
example of what Henri {\sc Poincar\'e} has called a \textit{limit cycle}
\cite{Andro1}, because integral curves approach it asymptotically.''
\end{quote}
Nevertheless, he quoted the paper in which Elie Cartan (1869-1951) and his son
Henri Cartan (1904-2008) established the uniqueness of the periodic solution to
Janet's equation (\ref{Jan1}) \cite{Ca}.

\begin{figure}[ht]
  \begin{center}
    \includegraphics[height=1.90cm]{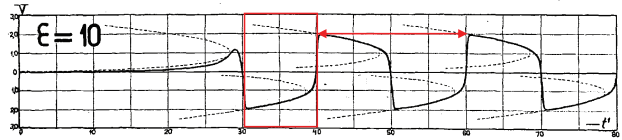} \\[-0.3cm]
    \caption{Relaxation oscillations solution to the van der Pol equation for
$\epsilon = 10$. From \cite[p. 986]{VdP26}.}
    \label{Vdp2}
  \end{center}
\end{figure}

Van der Pol \cite[p. 987]{VdP26} determined graphically the period of
relaxation oscillations starting from the time series (Fig. \ref{Vdp2}).
According to this figure, it is obvious that the period for relaxation
oscillations (from a maximum to the next maximum) is equal to twenty and not
approximatively equal to $\varepsilon = 10$ as claimed by van der Pol
\cite[p. 987]{VdP26}. It is rather surprising that van der Pol did not provide
$T = 2 \epsilon$, particularly because he correctly estimated the period to
be $2 \pi$ for $\epsilon = 0.1$. This lack of rigor was pointed out by Alfred
Li\'enard (1869-1958) in \cite[p. 952]{Lien} where he demonstrated the
existence and uniqueness of the periodic solution to a generalized van der Pol
equation.  Van der Pol had already realized his mistake and provided (in 1927)
a better approximation for the period according to the formula
\cite[p. 114]{VdP7}:
\begin{equation}
  \label{eqV4}
  T = \left( {3-2\log _e 2} \right)\varepsilon = 1.61\varepsilon \, .
\end{equation}

For $\epsilon = 10$, this leads to $T=16.1$ which
is still far away from the measured value of 20. Seventeen years later,
Jules Haag (1882-1953) provided a better approximation \cite{Ha1} by using
\textit{Asymptotic Methods} \cite[p. 103]{Ha2}:
\begin{equation}
  \label{eqH1}
  \begin{array}{rl}
    T & =  \displaystyle
     \left[ \displaystyle 3 - 2 \log \left( 2 \right) \right] \varepsilon
    + \frac{12.89}{\varepsilon^{1/3}} \\[0.3cm]
    & \displaystyle  ~~~
    + \frac{2}{\varepsilon}\left[ {-3.31+ \frac{19}{9} \log
    \left( \varepsilon \right)} \right]
    - \frac{4}{\varepsilon^{5/3}} + \cdots
  \end{array}
\end{equation}
Then, replacing $\varepsilon$ by ten into Eq. (\ref{eqH1}) one finds a value
(20.3) closer to twenty,.

\section{Turning relaxation oscillations as a concept}
\label{relos}

\subsection{Some insights on the German school}

German scientists were not quoted up to now. We will focus our attention on a
few contributions. Of course, sustained oscillations were found in a symmetrical
pendulum by Georg Duffing \cite{Duf18}. Sustained oscillations were found in a
thermionic system by Heinrich Barkhausen (1881-1956) and Karl Kurz
\cite{Bar19}.
It seems that this contribution triggered the interest of German physicists
for sustained oscillations in electro-technical or electronic devices. An
important contribution was provided by Erich Friedländer in 1926, a few months
before van der Pol's papers were published. Note that Friedl\"ander is quite
often quoted in \cite{Andro2}. Entitled ``On relaxation
oscillations ({\it Kippschwingungen}) in electronic vacuum tubes'', Friedländer
started his paper as follows \cite{Fri26a}.

\begin{figure}[ht]
  \begin{center}
    \begin{tabular}{cc}
      \includegraphics[height=2.4cm]{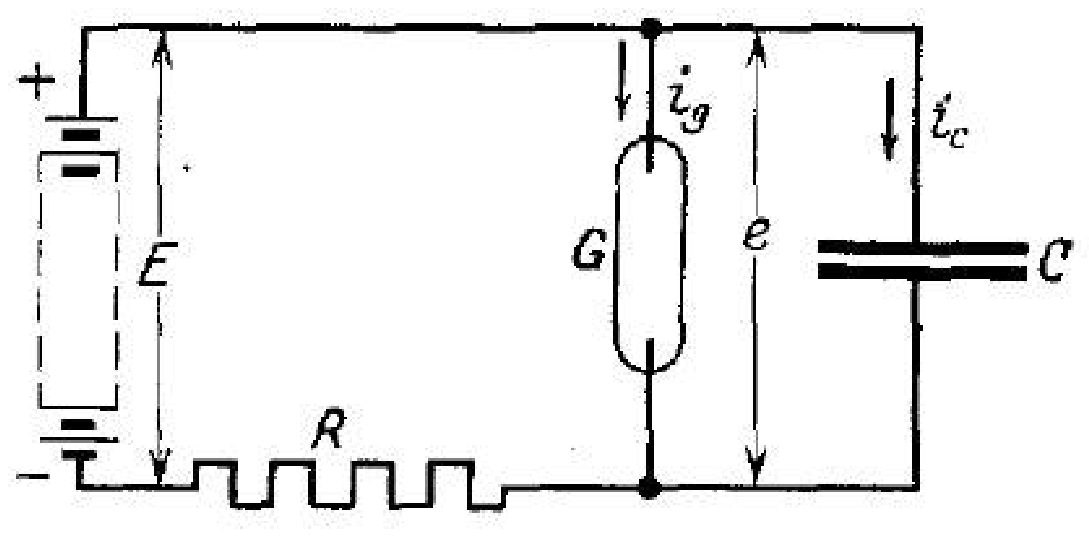} &
      \includegraphics[height=2.4cm]{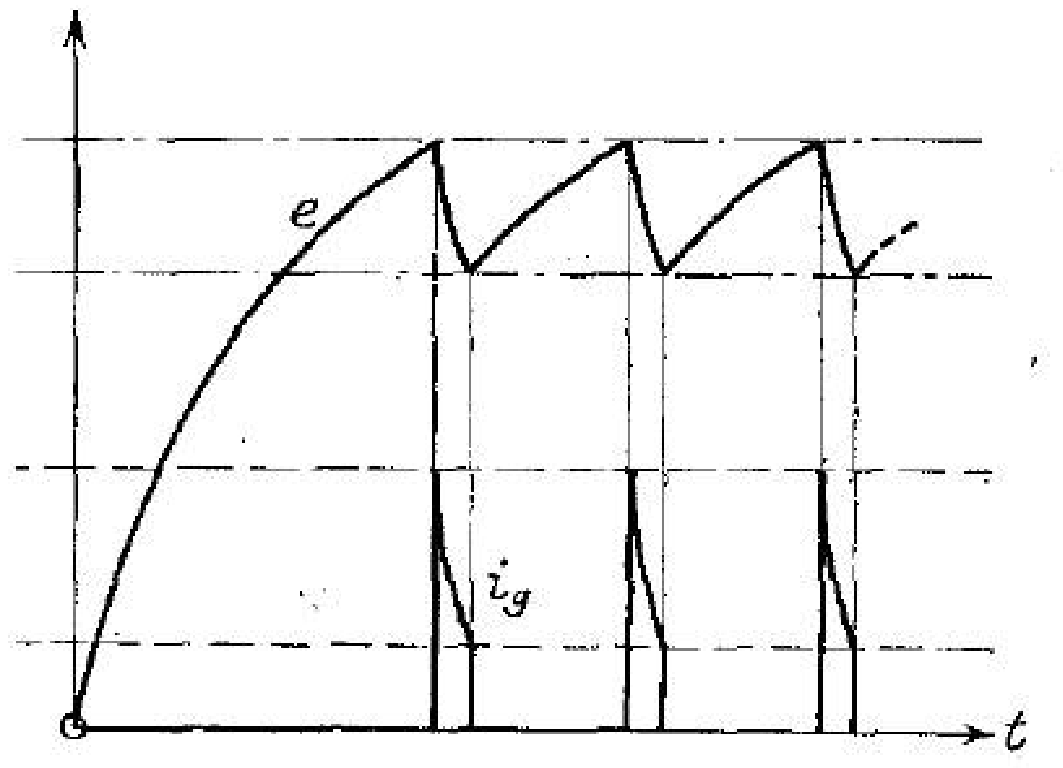} \\
      (a) Block diagram & (b) Nonlinear oscillations \\[-0.2cm]
    \end{tabular}
    \caption{Oscillations in a glow discharge circuit. From \cite{Fri26a}.}
    \label{glodis}
  \end{center}
\end{figure}

\begin{quote}
``it has been already pointed out that several systems can produce self-excited
oscillations which do not obey to the Thomson formula. [...] The best control
devices which are enable to enforce such an energy oscillating exchange are now
arc, neon lamp, vacuum tube.''
\end{quote}
Friedländer then evidenced an oscillating regime in a glow discharge
(Fig.\ \ref{glodis}).

\begin{figure}[ht]
  \begin{center}
    \begin{tabular}{cc}
      \includegraphics[height=2.8cm]{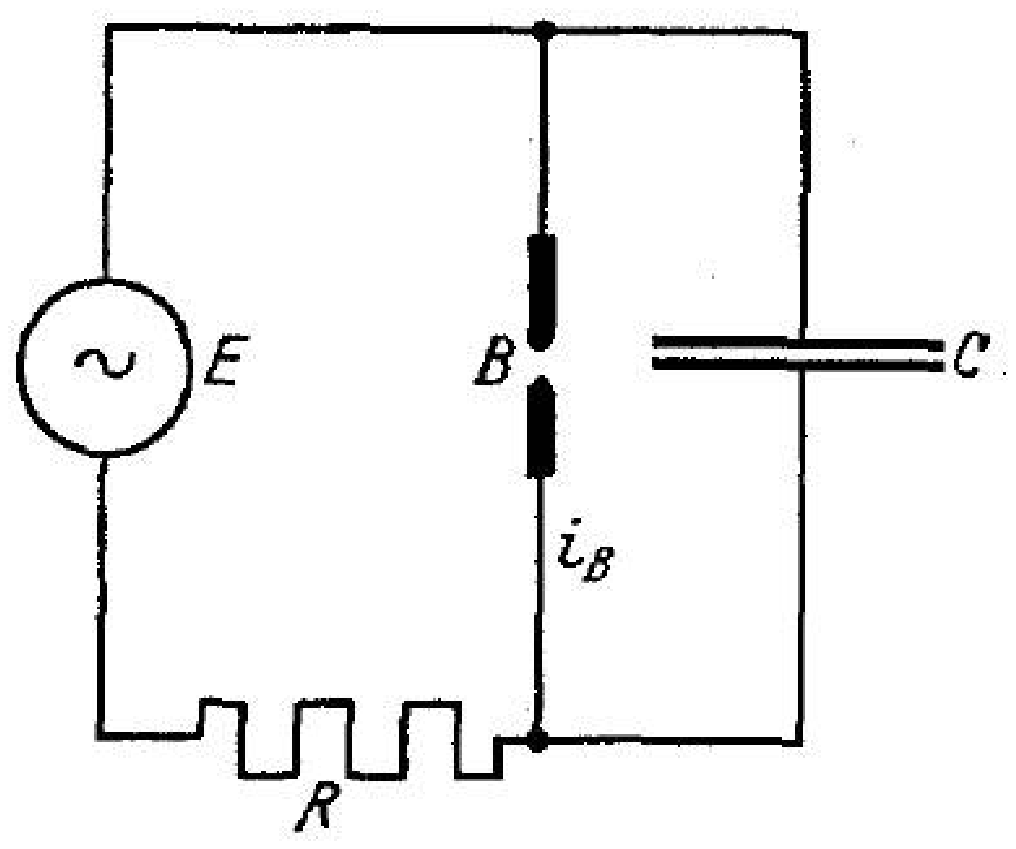} &
      \includegraphics[height=2.8cm]{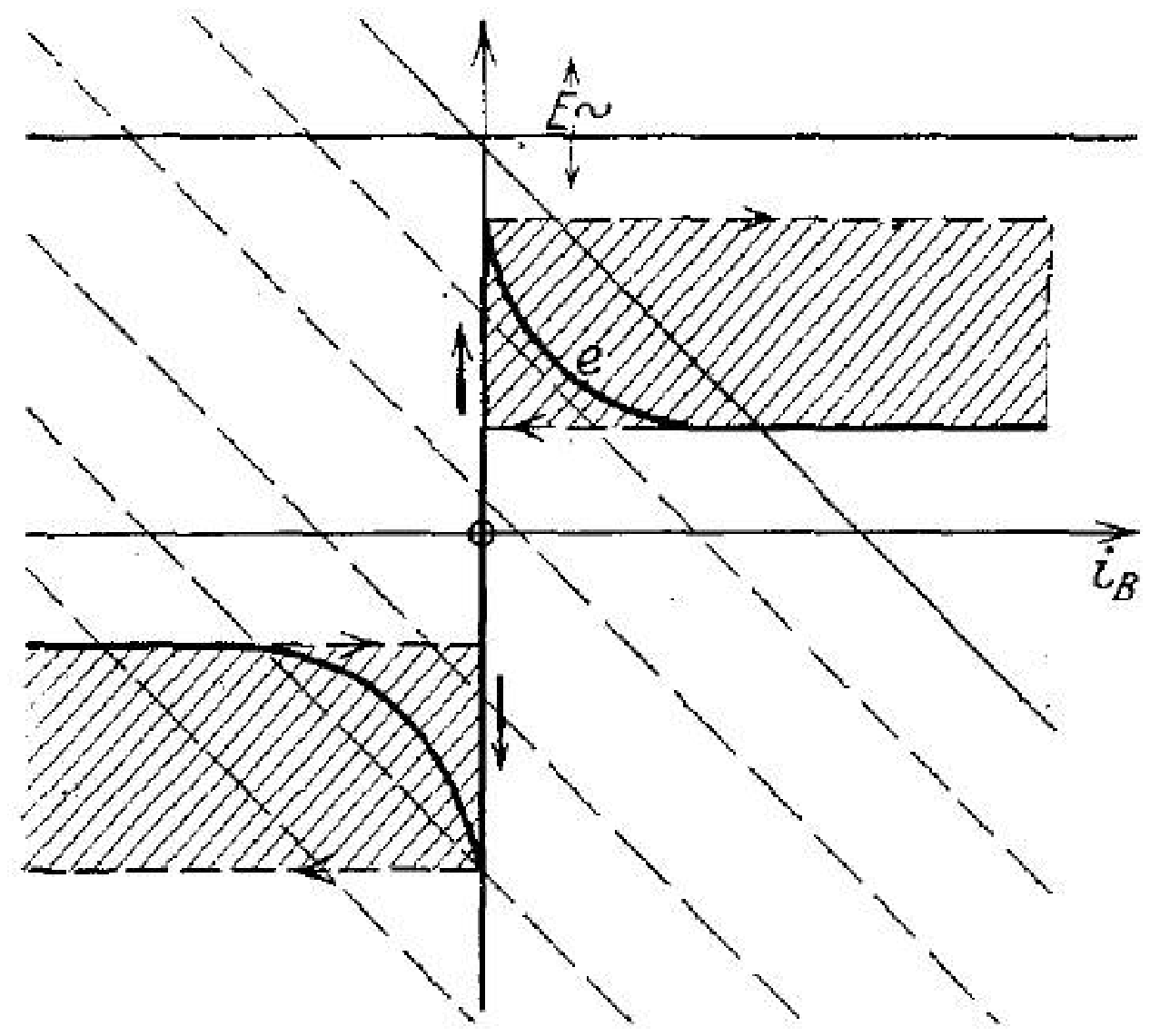} \\
      (a) Block diagram & (b) Circulation diagram \\[0.2cm]
      \multicolumn{2}{c}{
        \includegraphics[height=3.0cm]{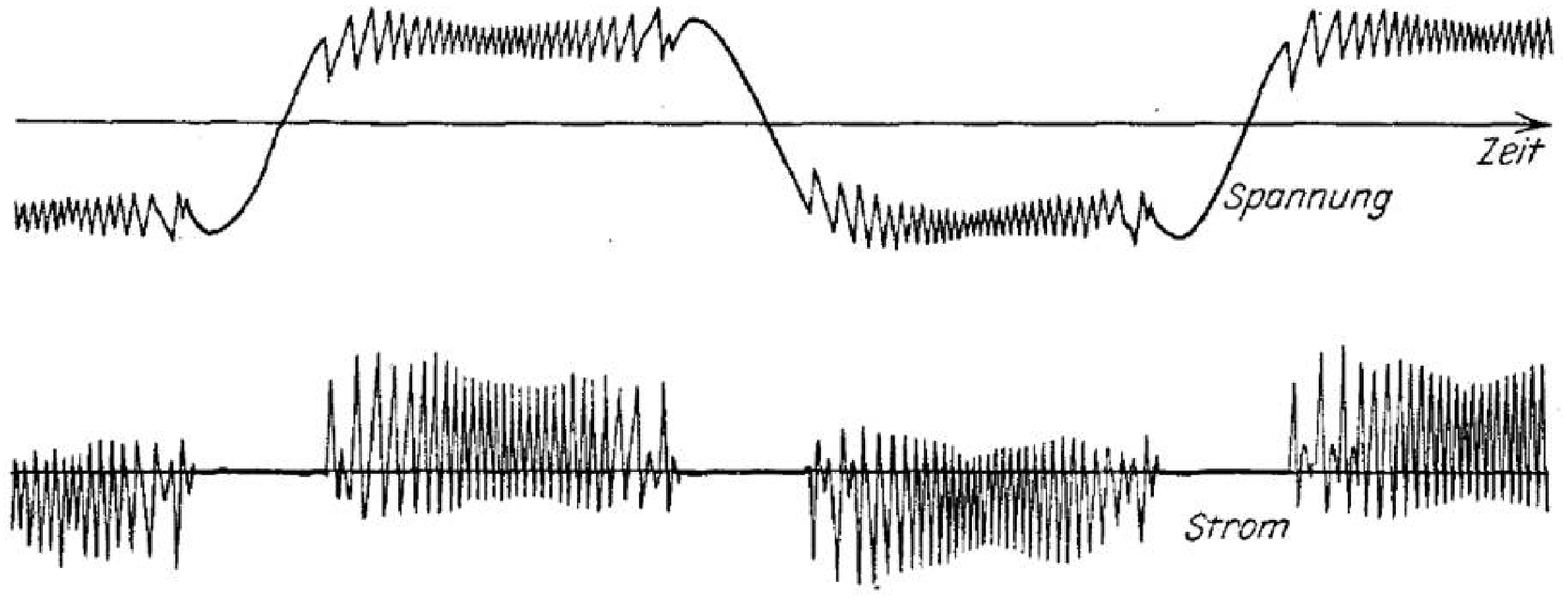}} \\
      \multicolumn{2}{c}{(c) Intermittent oscillations} \\[-0.2cm]
    \end{tabular}
    \caption{Oscillations in a glow discharge circuit. From \cite{Fri26a}.}
    \label{arclam}
  \end{center}
\end{figure}

Friedländer then explained how an alternating potential (Fig.\ \ref{arclam}b)
occurs in a circuit with  an arc lamp (Fig.\ \ref{arclam}a). With such a
circuit, he obtained some intermittent regimes as shown in Fig.\ \ref{arclam}c,
and which correspond to Duddell's intermittent regime or Blondel's
discontinuous arc.

\begin{figure}[ht]
  \begin{center}
    \begin{tabular}{cc}
      \includegraphics[height=3.8cm]{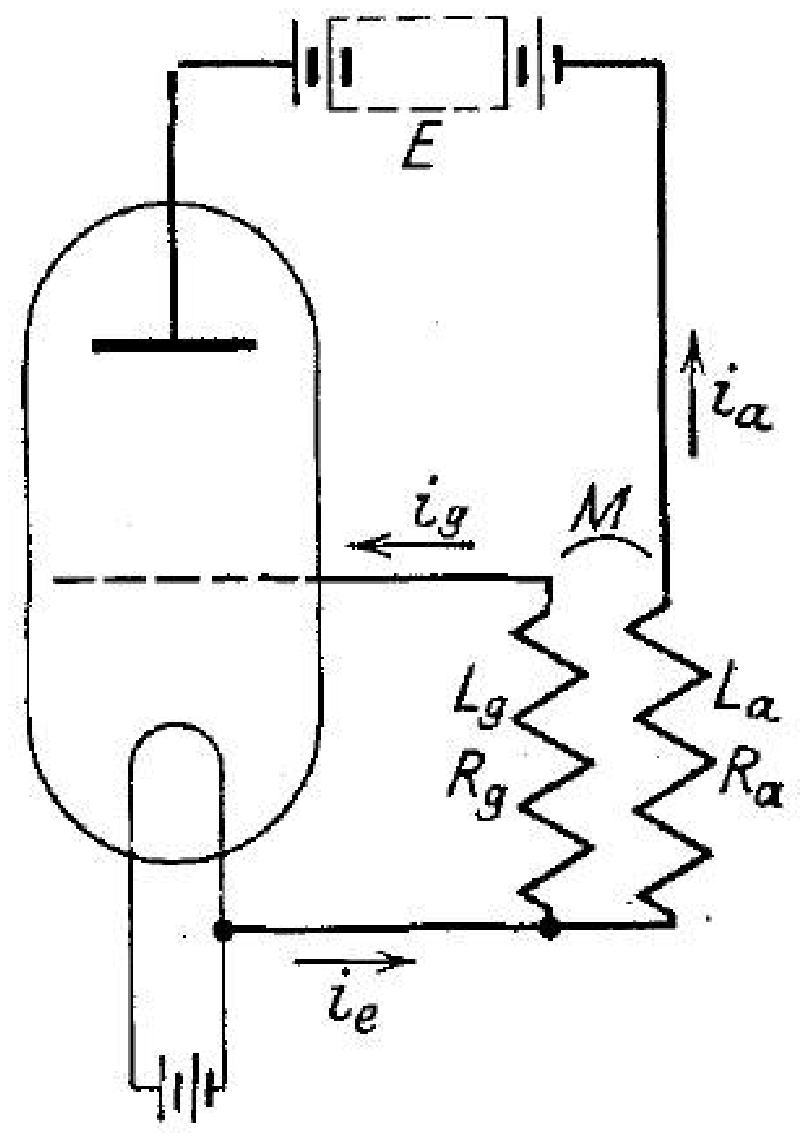} &
      \includegraphics[height=4.2cm]{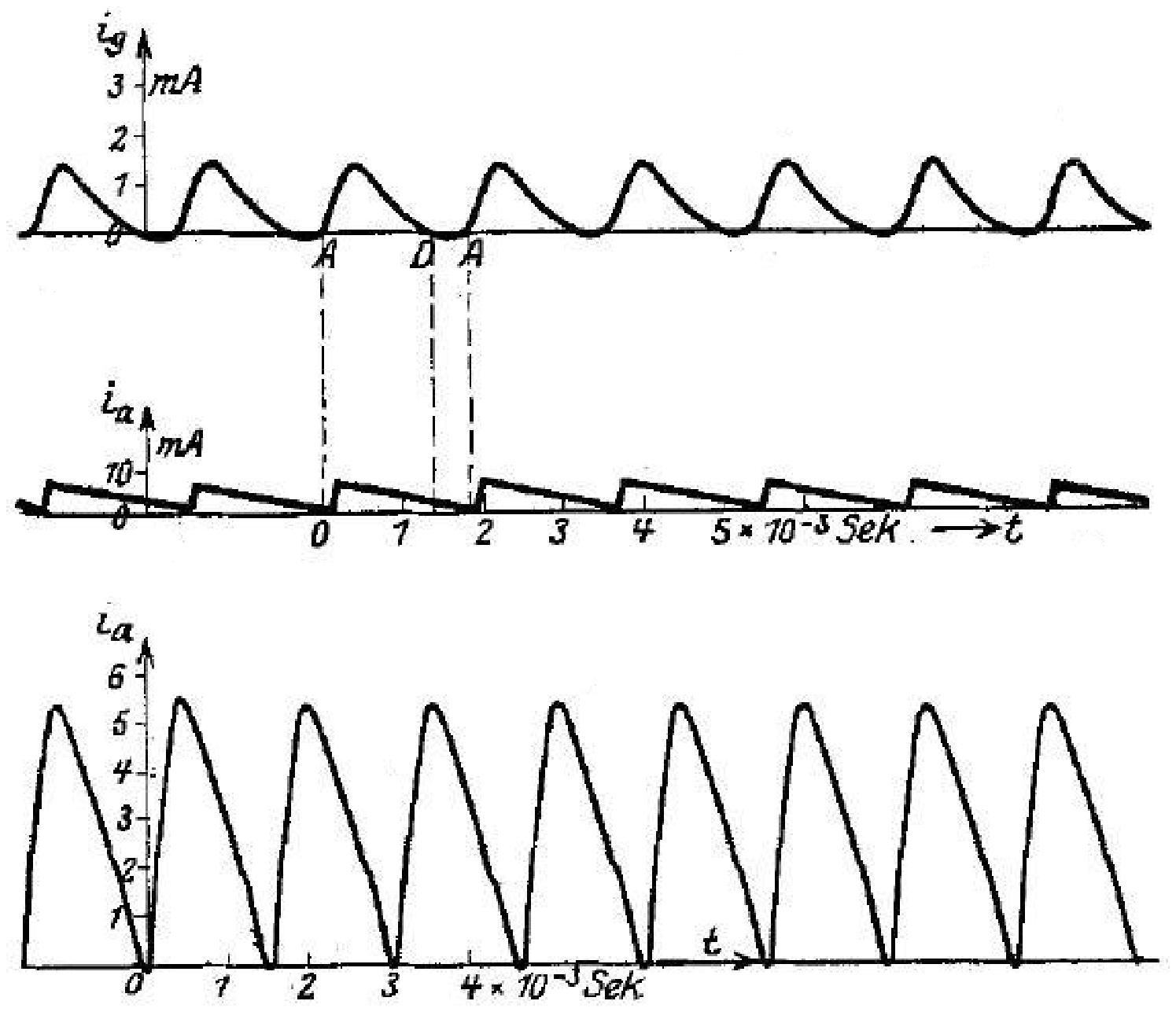} \\
      (a) Block diagram & (b) Relaxation oscillations \\[-0.2cm]
    \end{tabular}
    \caption{Sustained oscillations in a circuit with a triode. From
\cite{Fri26a}.}
    \label{tricic}
  \end{center}
\end{figure}

Friedländer also investigated a triode circuit (Fig.\
\ref{tricic}a) where he observed relaxation oscillations (Fig.\ \ref{tricic}b)
when the grid current did not reach saturation. He then investigated stability
conditions for some particular regimes  in a second paper \cite{Fri26b}.

\begin{figure}[ht]
  \begin{center}
    \includegraphics[height=4.5cm]{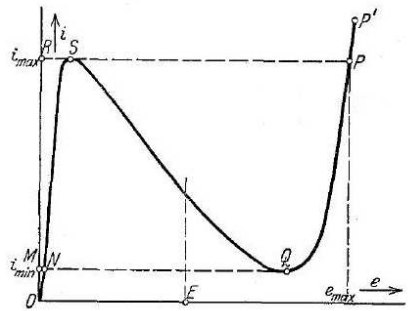} \\[-0.4cm]
    \caption{Cubic characteristic of the triode circuit investigated by Hudec
\cite{Hud29}.}
    \label{cubic}
  \end{center}
\end{figure}

\begin{figure}[ht]
  \begin{center}
    \includegraphics[height=5.7cm]{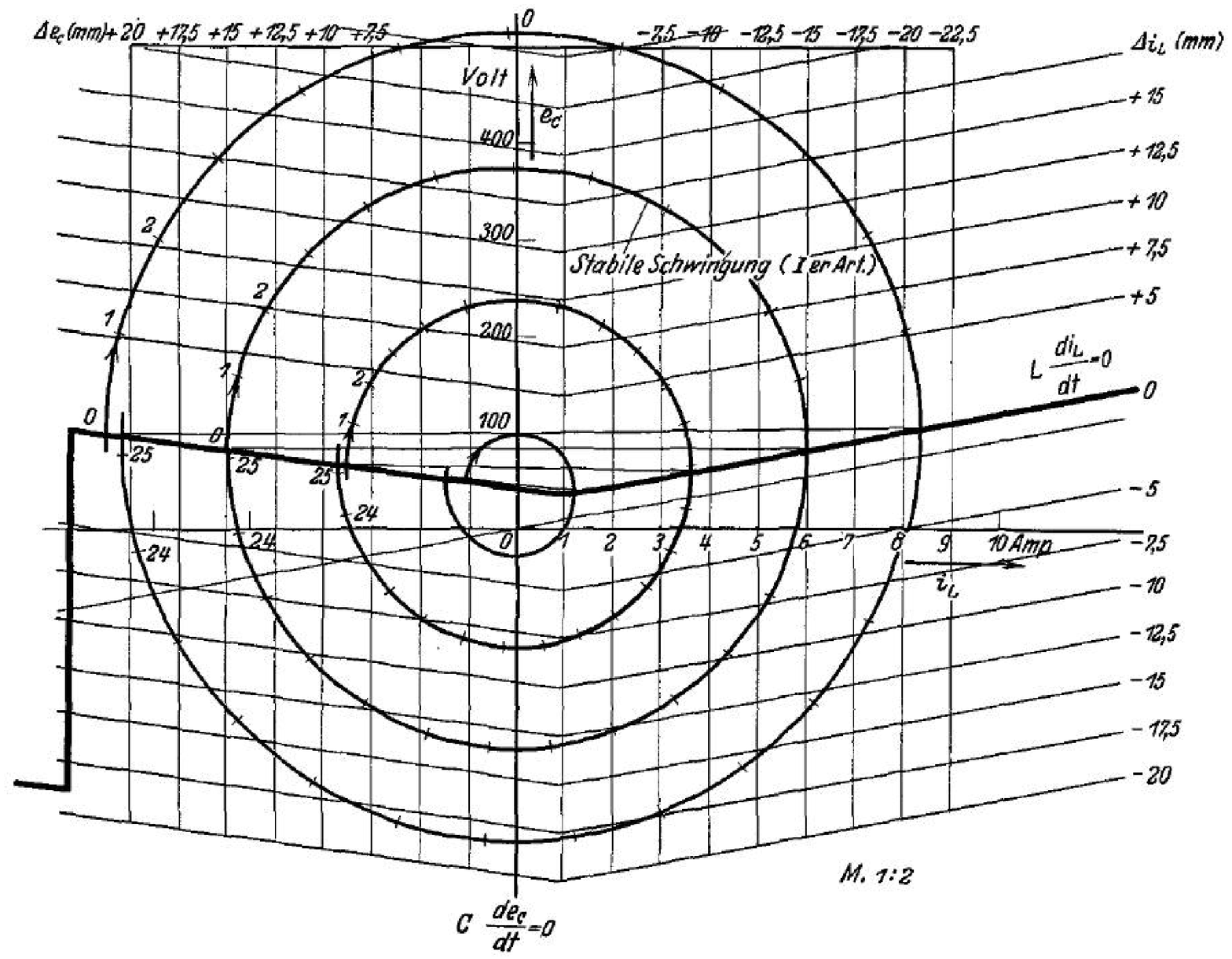} \\[-0.3cm]
    \caption{Plane diagram for representing solution to the triode circuit
as investigated by Kirschstein \cite{Kir30}.}
    \label{stasus}
  \end{center}
\end{figure}

Few years later Erich Hudec developed a theorie for ``Kippschwingungen'' in a
triode circuit with a cubic characteristic (Fig.\ \ref{cubic}), and presenting
various relaxation oscillations \cite{Hud29}. Hudec quoted Friedländer but
not van der Pol. Friedrich Kirschstein presented many relaxation oscillations
observed in various electronic circuits \cite{Kir30}. Contrary to Hudec,
he prefered the word ``Relaxationsschwingungen'' to ``Kippschwingungen'', and
quoted van der Pol \cite{VdP5}. With the graphical method used by van der Pol
(Fig.\ \ref{solnull}), he was looking for stable sustained solution (Fig.\
\ref{stasus}), but as in van der Pol's work, no link was made with
Poincar\'e's limit cycle. To end this brief review, W. Pupp did a similar
contribution one year before Kirschstein, showing stable sustained
oscillations --- in a plane projection --- produced by an impulse radio
emitter \cite{Pup29}.

\subsection{Van der Pol's lectures in France}

In 1926, van der Pol described the relaxation-oscillations as follows
\cite{VdP26}.

\begin{quote}
``It is seen that the amplitude alters very slowly from the value 2 to the
value 1 and then very suddenly it drops to the value -2. Next we observe
 a very gradual increase from the value -2 to the value -1 and again a
sudden jump to the value 2. This cycle term proceeds indefinitely.''
\end{quote}
As van der Pol later precised \cite{VdP8}

\begin{quote}
``their form [...] is characterized by discontinuous jumps arising each time
the system becomes unstable, that is, periodically [and] their period is
determined by a relaxation time.''
\end{quote}
He also remarked that these oscillations are easily synchronized with a
periodic external forcing term \cite{vdP27}. In 1931, Le Corbeiller described
them as ``oscillations whose shape is very different and whose prototype is
provided by a slow charge of a capacitor, followed by its rapid discharge, the
cycle being indefinitely repeated'' \cite{LeC33}.

As we showed the French school of electrical engineers and/or mathematicians
contributed quite a lot to the early developments of a theory in non-linear
oscillations. As briefly mentionned in the previous section, German school
was active too, but it was a little bit isolated, probably due to the
particular
political context during this period. Its actual contribution remains to be
more extensively investigated. In anyway, it is not surprising that van der
Pol, deeply involved in this area of research, was regularly invited in France
to present his works on relaxation oscillations. During each of his lectures
he exhibited the differential equation (\ref{protovdp}). Then, van
der Pol \cite[p. 371]{VdP8} provided a non-exhaustive list of examples:

\begin{quote}
``Hence with Aeolian harp, as with the wind blowing against telegraph wires,
causing a whistling sound, the time period of the sound heard is determined by
a diffusion-or relaxation time and has nothing to do with the natural period of
the string when oscillating in a sinusoidal way. Many other instances of
relaxation oscillations can be cited, such as: a pneumatic hammer, the
scratching noise of a knife on a plate, the waving of a flag in the wind, the
humming noise sometimes made by a watertap, the squeaking of a door, a steam
engine with much too small flywheel, [...], the periodic reoccurence of
epidemics and economical crises, the periodic density of an even number of
species of animals, living together and the one species serving as food to the
other, the sleeping of flowers, the periodic reoccurence of showers behind a
depression, the shivering from cold, the menstruation and finally the beating of
the heart.''
\end{quote}
Starting from these examples, van der Pol wanted to show that relaxation
oscillations are ubiquitous in nature. All of them, although apparently very
different, belong in fact to the new class of relaxation oscillations he
discovered as solutions to ``his'' prototypical equation (\ref{protovdp}).

With such an attitude, van der Pol tried to generalize the notion of relaxation
oscillations in order to confer it the value of a concept. This attempt for a
generalization was received with a mixed feeling in France.
On the one hand, a certain scepticism on behalf of the French and international
scientific community like that expressed by Li\'enard \cite[p. 952]{Lien} who
noted that the period of relaxation oscillations van der Pol graphically
deduced \cite{VdP26} was wrong or Rocard \cite[p. 402]{Ro} which was speaking
of a ``very extended generalization of the concept of relaxation oscillations''.
The French school was complaining about a certain lack of rigor in van der
Pol's works.
In U.S.S.R., Leonid Mandel'shtam \cite{Mandel} and his students
Andronov, Khaikin \& Witt \cite{Andro2} did not make use of the terminology
``relaxation oscillations'' introduced by van der Pol. They prefered the term
``discontinuous motions'', as used by Blondel, particularly because it
suggests a description of these oscillations in terms of slow/fast regimes.
This approach only became mature in the context of the singular perturbation
theory whose according to O'Malley \cite{Mal91}, early motivations arose from
Poincar\'e's works in celestial mechanics \cite{PoiMNC}, works by Ludwig
Prandtl (1875-1953) in fluid mechanics \cite{Pra05} and van der Pol's
contribution.

\begin{figure}[ht]
  \begin{center}
    \begin{tabular}{c}
      \includegraphics[width=6cm,height=1.6cm]{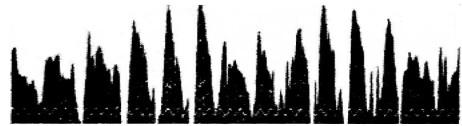} \\
      \begin{tabular}{c}
        \multicolumn{1}{p{7cm}}{(a) Global electromyogram of a strong voluntary
contraction by a men. }
      \end{tabular} \\[0.3cm]
      \includegraphics[height=0.9cm]{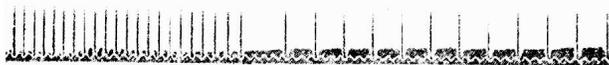} \\
      \begin{tabular}{c}
        \multicolumn{1}{p{7cm}}{(b) Pulsation of a nervous ending observed
using the local action potential.}
      \end{tabular} \\[-0.2cm]
    \end{tabular}
    \caption{Two physiological examples where relaxation oscillations were
found by Fessard \cite{Fes}.}
    \label{Fesfig}
  \end{center}
\end{figure}

\begin{figure}[ht]
  \begin{center}
    \begin{tabular}{c}
      \includegraphics[height=1.55cm]{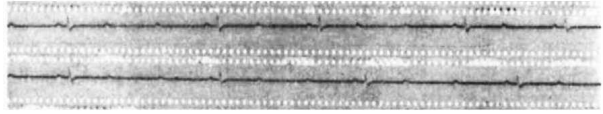} \\
      (a) Heartbeats \\[0.2cm]
      \includegraphics[height=3.8cm]{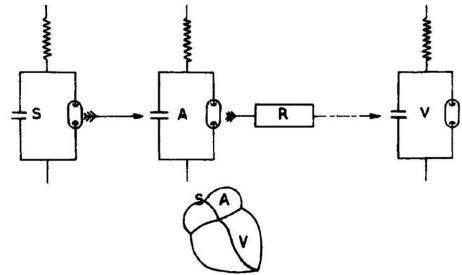} \\
      (b) The heart model as three three relaxation systems \\[-0.2cm]
    \end{tabular}
    \caption{Heartbeats produced by a heart model made of three relaxation
systems. S $\equiv$ sinus, A $\equiv$ auriculum and V $\equiv$ ventriculum.
R is a delay system for reproducing the time necessary for a stimulus to be
transmitted through the auriculo-ventricular bundle. From \cite{VdP40}.}
    \label{heart}
  \end{center}
\end{figure}

\begin{figure}[ht]
  \begin{center}
    \includegraphics[height=3.3cm]{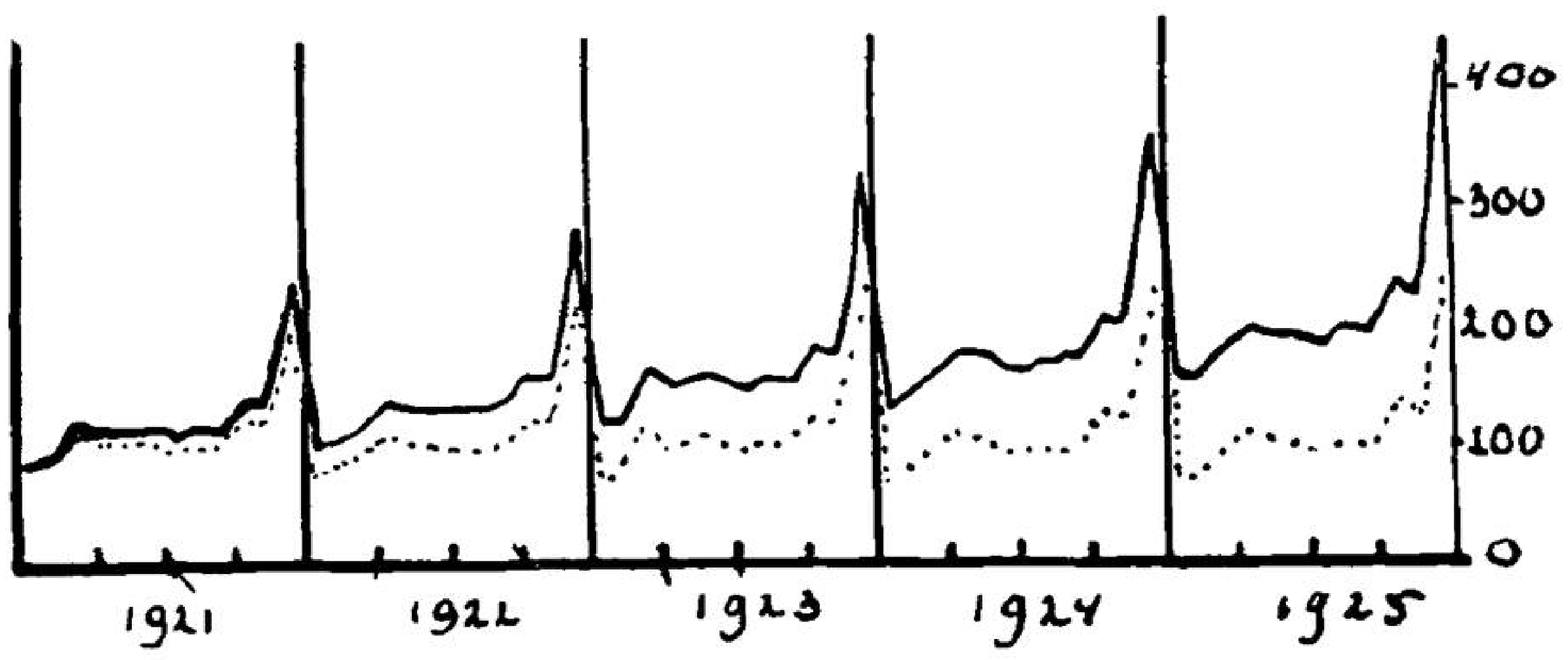} \\[-0.4cm]
    \caption{Ratios of monthly sales in shop ``Five-and-ten chain stores'' in
USA between 1921 and 1925. The solid line corresponds to the real data, the
dotted curve to what occurs after eliminating the annual growth. From Hamburger
\cite{Hamb}.}
    \label{Hamb1}
  \end{center}
\end{figure}

On the other hand, an extreme focusing on relaxation oscillations took
the shape of a ``hunting of the relaxation effect''. Minorsky spent the fourth
part of his book \cite{Min47} on relaxation oscillations, quoting various
examples as suggested in Le Corbeiller's book \cite{LeCorb}. In France, the
neurophysiologist Alfred Fessard (1900-1982) exhibited relaxation
oscillations in nervous rhythms (Fig.\ \ref{Fesfig}) \cite{Fes}, the engineer
in hydrodynamics
François-Joseph Bourrières (1880-1970) found some of them in a hose through
which a fluid was flowing \cite{Bou}. Concerning the list of typical examples
proposed by van der Pol \cite[p. 371]{VdP8} some of them could be considered as
being non justified. But van der Pol \cite{VdP40} showed quite
convincing time series (Fig.\ \ref{heart}a) produced by a heart model made of
three relaxation systems (Fig.\ \ref{heart}b). Some periodic reoccurence of
economical crisis were studied at that time by the Dutch economist
Ludwig Hamburger \cite{Hamb}. He explained that sales curves correspond in many
aspects to relaxation oscillations (Fig. \ref{Hamb1}).

\begin{figure}[ht]
  \begin{center}
    \includegraphics[height=4.5cm]{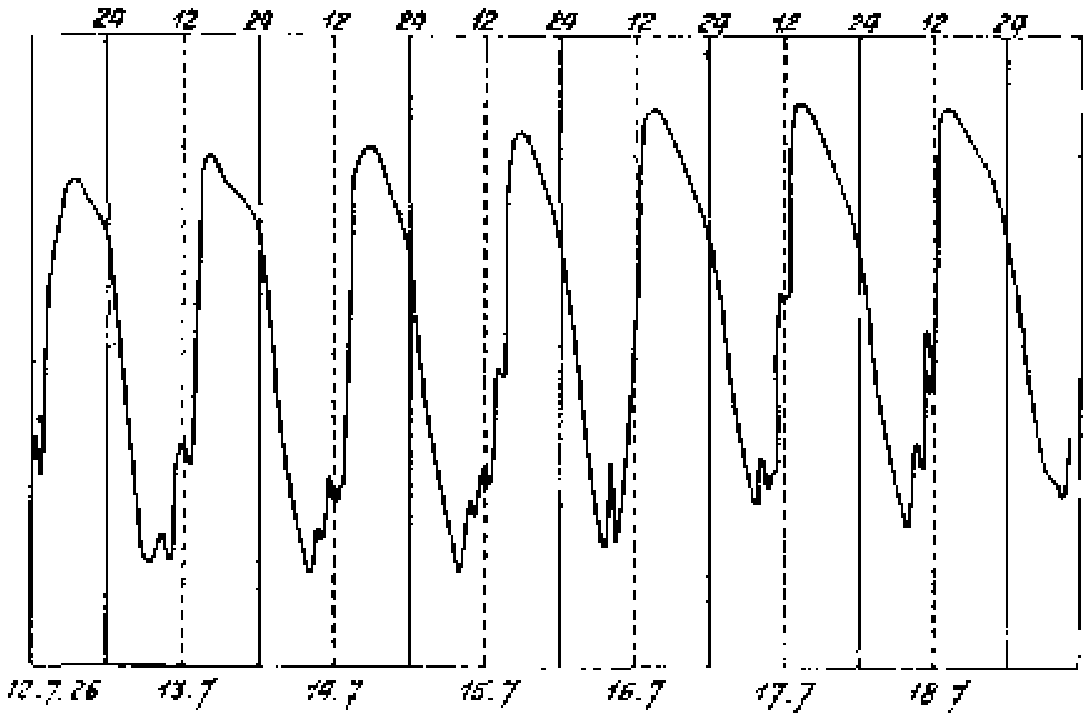} \\[-0.2cm]
    \caption{Leaf movement of {\it Canavalia ensiformis} plants revealing a
circadian rhythm and considered as a relaxation oscillation by van der Pol.
From Kleinhoonte \cite{Klein}.}
    \label{Klein1}
  \end{center}
\end{figure}

Sleep of flowers was also seriously investigated by Antonia Kleinhoonte
in her thesis \cite{Klein}. The leaf movement of jack-bean, {\it Canavalia
ensiformis}, plants daily fluctuates as relaxation oscillations. Kleinhoonte's
work is considered as a confirmation of a circadian rhythm in plants, as
proposed by Jagadish Chandra Bose \cite{Bose}. The time series shown in Fig.
\ref{Klein1} was included in one of van der Pol's paper \cite[p. 312]{VdP9}.
Nonlinear oscillations were already observed in the Lotka-Volterra model
\cite{Lot20,Vol26}; although this model was conservative, the obtained
oscillations were nonlinear and do not look too different from the relaxation
oscillations observed in dissipative systems.

\subsection{Le Corbeiller's contribution}

In the early 1930s, the French engineer Philippe Le Corbeiller (1891-1930),
who was assisting van der Pol during his lectures in France, contributed to
popularize the concept of relaxation oscillations. After the First World War,
he became an expert in electronics, acoustics and worked for the French
Ministry of Communication. In his lectures, he most often promoted van der
Pol's contribution \cite[p. 4]{LeCorb} as for instance follows.

\begin{quote}
`` [...] this is a Dutch physicist, M. Balth. van der Pol, who, by his theory
on \textit{relaxation oscillations} (1926) made decisive advances. Scholars
from various countries are working today to expand the path he traced; among
these contributions, the most important seems to be that of M. Li\'enard
(1928). Very interesting mathematical research are pursued by M. Andronow from
Moscow.''
\end{quote}
Poincar\'e was mentioned for the limit cycle that Andronov identified
in 1929 \cite{Andro1}. Blondel was quoted for having suggested the term of
``variance'' for negative resistance, Janet for his popularization of the
G\'erard-Lescuyer experiments, but Duddell was forgotten. With such a
presentation, van der Pol was presented as the main contributor who triggered
the interesting approach (before was the ``old'' science, after was
the ``modern'' science). This is in fact only based on the word ``relaxation''.
When he spoke about preceding works, only the experimental evidences supporting
van der Pol's relaxation theory were mentionned.
In his conclusion, Le Corbeiller \cite[p. 45]{LeCorb} seems to come back to a
more realistic point of view:

\begin{quote}
``The mathematical theory of relaxation oscillations only begun, several
phenomena of which we spoke are just empirically known and, therefore, can
surprise us [{\ldots}]. In a word, we have in the front of us the immense field
of \textit{nonlinear systems}, in which we have only begun to penetrate. I would
hope that these conferences have shown all the interest we can have in its
exploration.''
\end{quote}
But the story was already written: Balthazar van der Pol was the first one to
give a ``general framework'' to investigate all these behaviors. He was thus
promoted as the heroe of this field. To be complete about Le Corbeiller's
position, let us
mention that he became close enough to van der Pol's widow Pietronetta
Posthuma (1893-1984) to marry her in New York, on May 7, 1964.

\section{Conclusion}
\label{conc}

We showed that many self-sustained oscillating experiments were observed before
the 1920s, some of them being well investigated, experimentally as well as
theoretically. Blondel should be mentionned as being the one who pushed quite
far explainations for the physical mechanisms underlying such phenomena. The
key property of these new oscillations was that they were not linear, that is,
not associated with a frequency matching with the eigen-frequency of the
circuit (Thomson's formula). Blondel noted such a property. Van der Pol
then distinguished a particular type of oscillations which greatly differ from
sinusoidal oscilaltions and whose period does not match with the Thomson
formula. He proposed a generic name, the ``relaxation oscillations'', but no
clear mathematical definition was proposed.  A ``slow evolution followed by a
sudden jump'' is not enough to avoid ambiguous interpretation. As a
consequence, many ``relaxation oscillations'' were seen in various
contexts. If we consider that relaxation is justified by a kind of ``discharge
property'' by analogy with a capacitor, some of the examples provided by
van der Pol, clearly not belong to this class of oscillations, as well
exemplified by the plant leaf movement.

The so-called van der Pol equation --- in its interesting simplified form ---
was written for the first time by van der Pol. This equation was not the first
one where self-oscillating behaviours were observed but this is the first
``reduced equation'' in Curie's sense. The preceding equations were obtained
for some particular experimental set-up as exemplified by Poincar\'e, Blondel
and Janet. Van der Pol spent a large amount of energy to convince people that
his simple equation and its solutions could serve as a paradigm to explain
various behaviors: the breakthrough was also that what is important is the
dynamics and no longer the physical processes. With its graphical
representation, he also contributed to change the way in which these nonlinear
phenomena had to be investigated. To that respect, associating
his name with this simple equation is one of the possibilities to thank him.

\section{Remark}

This article has been entirely copied by C. Letellier in the chapter 2 pp. 101-137 of his book entitled \textit{Chaos in Nature} but in omitting the reference to the present work which was directly issued from the original research made by Ginoux \cite{Gith} in his second PhD-thesis in \textit{History of Sciences}.

\end{document}